\documentclass[ALICE,manyauthors]{cernphprep}
\usepackage[comma,square,numbers,sort&compress]{natbib}
 \usepackage{xspace}
 \usepackage{hyperref}
 \usepackage[T1]{fontenc}

 \usepackage{color}
 \graphicspath{{figure/}}

\begin{document}
%

\newcommand{\pp}           {pp\xspace}
\newcommand{\ppbar}        {\mbox{$\mathrm {p\overline{p}}$}\xspace}
\newcommand{\XeXe}         {\mbox{Xe--Xe}\xspace}
\newcommand{\PbPb}         {\mbox{Pb--Pb}\xspace}
\newcommand{\pA}           {\mbox{pA}\xspace}
\newcommand{\pPb}          {\mbox{p--Pb}\xspace}
\newcommand{\AuAu}         {\mbox{Au--Au}\xspace}
\newcommand{\dAu}          {\mbox{d--Au}\xspace}

\newcommand{\s}            {\ensuremath{\sqrt{s}}\xspace}
\newcommand{\snn}          {\ensuremath{\sqrt{s_{\mathrm{NN}}}}\xspace}
\newcommand{\pt}           {\ensuremath{p_{\rm T}}\xspace}
\newcommand{\pT}           {\ensuremath{p_{\rm T}}\xspace}
\newcommand{\meanpt}       {$\langle p_{\mathrm{T}}\rangle$\xspace}
\newcommand{\ycms}         {\ensuremath{y_{\rm CMS}}\xspace}
\newcommand{\ylab}         {\ensuremath{y_{\rm lab}}\xspace}
\newcommand{\etarange}[1]  {\mbox{$\left | \eta \right |~<~#1$}}
\newcommand{\yrange}[1]    {\mbox{$\left | y \right |~<~#1$}}
\newcommand{\dndy}         {\ensuremath{\mathrm{d}N_\mathrm{ch}/\mathrm{d}y}\xspace}
\newcommand{\dndeta}       {\ensuremath{\mathrm{d}N_\mathrm{ch}/\mathrm{d}\eta}\xspace}
\newcommand{\avdndeta}     {\ensuremath{\langle\dndeta\rangle}\xspace}
\newcommand{\dNdy}         {\ensuremath{\mathrm{d}N_\mathrm{ch}/\mathrm{d}y}\xspace}
\newcommand{\Npart}        {\ensuremath{N_\mathrm{part}}\xspace}
\newcommand{\Ncoll}        {\ensuremath{N_\mathrm{coll}}\xspace}
\newcommand{\dEdx}         {\ensuremath{\textrm{d}E/\textrm{d}x}\xspace}
\newcommand{\RpPb}         {\ensuremath{R_{\rm pPb}}\xspace}
\newcommand{\kT}            {$k_{T}$}
\newcommand{\deltaeta}      {$\Delta \eta$}
\newcommand{\aveNch}    {$\langle N_{\rm ch} \rangle$}
\newcommand{\aveNacc}        {$\langle N_{\rm acc} \rangle$}
\newcommand{\aveN}          {$\langle N \rangle$}

\newcommand{\nineH}        {$\sqrt{s}~=~0.9$~Te\kern-.1emV\xspace}
\newcommand{\seven}        {$\sqrt{s}~=~7$~Te\kern-.1emV\xspace}
\newcommand{\twoH}         {$\sqrt{s}~=~0.2$~Te\kern-.1emV\xspace}
\newcommand{\twosevensix}  {$\sqrt{s}~=~2.76$~Te\kern-.1emV\xspace}
\newcommand{\five}         {$\sqrt{s}~=~5.02$~Te\kern-.1emV\xspace}
\newcommand{\twosevensixnn}{$\sqrt{s_{\mathrm{NN}}}~=~2.76$~Te\kern-.1emV\xspace}
\newcommand{\fivenn}       {$\sqrt{s_{\mathrm{NN}}}~=~5.02$~Te\kern-.1emV\xspace}
\newcommand{\LT}           {L{\'e}vy-Tsallis\xspace}
\newcommand{\GeVc}         {Ge\kern-.1emV/$c$\xspace}
\newcommand{\MeVc}         {Me\kern-.1emV/$c$\xspace}
\newcommand{\TeV}          {Te\kern-.1emV\xspace}
\newcommand{\GeV}          {Ge\kern-.1emV\xspace}
\newcommand{\MeV}          {Me\kern-.1emV\xspace}
\newcommand{\GeVmass}      {Ge\kern-.2emV/$c^2$\xspace}
\newcommand{\MeVmass}      {Me\kern-.2emV/$c^2$\xspace}
\newcommand{\lumi}         {\ensuremath{\mathcal{L}}\xspace}
\newcommand{\aveNpart} {$\langle N_{\rm part}\rangle$}

\newcommand{\ITS}          {\rm{ITS}\xspace}
\newcommand{\TOF}          {\rm{TOF}\xspace}
\newcommand{\ZDC}          {\rm{ZDC}\xspace}
\newcommand{\ZDCs}         {\rm{ZDCs}\xspace}
\newcommand{\ZNA}          {\rm{ZNA}\xspace}
\newcommand{\ZNC}          {\rm{ZNC}\xspace}
\newcommand{\SPD}          {\rm{SPD}\xspace}
\newcommand{\SDD}          {\rm{SDD}\xspace}
\newcommand{\SSD}          {\rm{SSD}\xspace}
\newcommand{\TPC}          {\rm{TPC}\xspace}
\newcommand{\TRD}          {\rm{TRD}\xspace}
\newcommand{\VZERO}        {\rm{V0}\xspace}
\newcommand{\VZEROA}       {\rm{V0A}\xspace}
\newcommand{\VZEROC}       {\rm{V0C}\xspace}
\newcommand{\Vdecay} 	   {\ensuremath{V^{0}}\xspace}

\newcommand{\ee}           {\ensuremath{e^{+}e^{-}}} 
\newcommand{\pip}          {\ensuremath{\pi^{+}}\xspace}
\newcommand{\pim}          {\ensuremath{\pi^{-}}\xspace}
\newcommand{\kap}          {\ensuremath{\rm{K}^{+}}\xspace}
\newcommand{\kam}          {\ensuremath{\rm{K}^{-}}\xspace}
\newcommand{\pbar}         {\ensuremath{\rm\overline{p}}\xspace}
\newcommand{\kzero}        {\ensuremath{{\rm K}^{0}_{\rm{S}}}\xspace}
\newcommand{\lmb}          {\ensuremath{\Lambda}\xspace}
\newcommand{\almb}         {\ensuremath{\overline{\Lambda}}\xspace}
\newcommand{\Om}           {\ensuremath{\Omega^-}\xspace}
\newcommand{\Mo}           {\ensuremath{\overline{\Omega}^+}\xspace}
\newcommand{\X}            {\ensuremath{\Xi^-}\xspace}
\newcommand{\Ix}           {\ensuremath{\overline{\Xi}^+}\xspace}
\newcommand{\Xis}          {\ensuremath{\Xi^{\pm}}\xspace}
\newcommand{\Oms}          {\ensuremath{\Omega^{\pm}}\xspace}
\newcommand{\degree}       {\ensuremath{^{\rm o}}\xspace}

 \begin{titlepage}
 \PHyear{2021}       
 \PHnumber{89}      
 \PHdate{11 May}  

 \title{Charged-particle multiplicity fluctuations in Pb--Pb collisions at \snn~=~2.76~TeV}   

 \Collaboration{ALICE Collaboration\thanks{See Appendix~\ref{app:collab} for the list of collaboration members}}
 \ShortAuthor{ALICE Collaboration} 

\begin{abstract}

Measurements of event-by-event fluctuations of charged-particle multiplicities in Pb--Pb collisions at \snn~$=$~2.76~TeV using the ALICE detector at the CERN Large Hadron Collider (LHC) are presented in the
pseudorapidity range $|\eta|<0.8$ and transverse momentum $0.2 < p_{\rm T} < 2.0$~GeV/$c$.
The amplitude of the fluctuations is expressed in terms of the variance normalized by the mean of the multiplicity distribution. The $\eta$ and $p_{\rm T}$ dependences of the fluctuations and their evolution with
respect to collision centrality are investigated. The multiplicity fluctuations tend to decrease from peripheral to central collisions. The results are compared to those obtained from HIJING and AMPT Monte Carlo event generators as well as to experimental data at lower collision energies. 
Additionally, the measured multiplicity fluctuations are discussed in the context of the isothermal compressibility of the high-density strongly-interacting system formed in central Pb--Pb collisions.

\end{abstract}
\end{titlepage}

\setcounter{page}{2} 


\section{Introduction}

According to quantum chromodynamics (QCD), at high temperatures and high energy densities, nuclear matter undergoes a phase transition to a deconfined state of quarks and gluons, the quark--gluon plasma (QGP)~\cite{Satz:2011wf,Ratti:2005jh,Bass:1998vz,Braun-Munzinger:2015hba,Andronic:2017pug}. 
Heavy-ion collisions at ultra-relativistic energies
make it possible to create and study such strongly-interacting matter under extreme conditions. The QGP formed in high-energy heavy-ion collisions has been characterised as a strongly-coupled system with very low shear viscosity. The primary goal of the heavy-ion program at the CERN Large Hadron
Collider (LHC) is to study the QCD phase structure by measuring the properties of QGP matter. One of the important methods for this study is the measurement of event-by-event fluctuations of experimental observables. These fluctuations are sensitive to the proximity of the phase transition and thus provide information on the nature and dynamics of the system formed in the collisions~\cite{Stephanov:1998dy,Stephanov:1999zu,Koch:2001zn,Jeon:2000wg,Jeon:2003gk,Karsch:2005ps,Stokic:2008jh}. Fluctuation measurements provide a powerful tool to investigate the response of a system to external perturbations. Theoretical developments suggest that it is possible to extract quantities related to the thermodynamic properties of the system, such as entropy, chemical potential, viscosity, specific heat, and
isothermal compressibility~\cite{VanHove:1983rk,Bialas:2002hm,Baym:1999up,Stephanov:1998dy,Shuryak:1997yj,Mrowczynski:1997kz,
Gazdzicki:2003bb,Mukherjee:2016yjq,Mukherjee:2017elm,Basu:2016ibk}. 
In particular, isothermal compressibility expresses how a system's volume responds to a change in the applied pressure. In the case of heavy-ion collisions, it has been shown that the isothermal compressibility can be calculated from the event-by-event fluctuation of charged-particle multiplicity distributions~\cite{Mrowczynski:1997kz}. 

The measured multiplicity scales with the collision centrality in heavy-ion collisions. The distribution of particle multiplicities in a given class of centrality and its fluctuations on an event-by-event basis provide information on particle production mechanisms~\cite{Heiselberg:2000fk,Begun:2004gs,Basu:2020jbk}. In this work, the magnitude of the fluctuations is quantified in terms of the scaled variance,
\begin{equation}
  \omega_{\rm ch} =\frac{\sigma_{\rm ch}^{\rm 2}}{\langle N_{\rm ch} \rangle},
\end{equation}
where $\langle N_{\rm ch} \rangle$ and $\sigma_{\rm ch}^2$~denote the mean and variance of the charged-particle 
multiplicity distribution, respectively.
Event-by-event multiplicity fluctuations in heavy-ion collisions have been studied earlier at
the BNL-AGS by E802~\cite{E-802:1995wgu}, 
the CERN-SPS by the WA98~\cite{Aggarwal:2001aa}, NA49~\cite{Alt:2006jr,Alt:2004ir}, and CERES~\cite{Sako:2004pw} experiments, and at the Relativistic Heavy Ion Collider (RHIC) by the PHOBOS~\cite{Wozniak:2004kp} and PHENIX~\cite{Adare:2008ns} experiments. A compilation of available experimental data and comparison to predictions of the event generators are presented elsewhere~\cite{Mukherjee:2016yjq}. In this work,  measurements of the scaled variance of multiplicity fluctuations are presented as a function of collision centrality in Pb--Pb collisions at \snn~$=$~2.76~TeV using the ALICE detector at the LHC.

In thermodynamics, the isothermal compressibility (\kT) is defined as the fractional change in the volume of a system with change of pressure at a constant temperature,
\begin{equation}
k_{T} =
-\frac{1}{V}\left.\left(\frac{\partial V}{\partial P}\right)\right|_{T},
\end{equation}
where $V,T,P$ are the volume, temperature, and pressure of the system, respectively. In general, an increase in the applied pressure leads to a decrease in volume, so the negative sign makes the value of \kT~positive.
In the context of a description in terms of the grand canonical ensemble, which is approximately applicable for the description of particle production in heavy-ion collisions~\cite{Andronic:2017pug}, the scaled variance of the multiplicity distribution can be expressed as~\cite{Mrowczynski:1997kz},
\begin{equation}
\omega_{\rm ch} = \frac{k_{\rm B}T \langle N_{\rm ch} \rangle}{V}k_{T},
\label{kTeqn}
\end{equation}
where $k_{\rm B}$~is the Boltzmann's constant, and $\langle N_{\rm ch}\rangle$  is the average 
number of charged particles. Measurements of fluctuations in terms of $\omega_{\rm ch}$
can be exploited to determine \kT~and associated thermodynamic quantities such as the speed of sound within the system~\cite{Mrowczynski:1997kz,Sahu:2020swd}.

Measurements of the multiplicity of produced particles in relativistic heavy-ion collisions
are basic to most of the studies as a majority of the experimentally observed quantities are directly related to the multiplicity. 
The variation of the multiplicity depends on the fluctuations in
the collision impact parameter or the number of participant nucleons.
Thus, the measured multiplicity fluctuations contain contributions from event-by-event fluctuations 
in the number of participant nucleons, which forms the main background towards the evaluation of any 
thermodynamic quantity~\cite{Begun:2017sgs,Braun-Munzinger:2016yjz}. This has been partly addressed by selecting narrow
intervals in centrality and accounting for the multiplicity variation within the centrality  
of the measurement. The remainder of participant fluctuations is estimated in the context of an MC Glauber model in which
nucleus--nucleus collisions are considered to be a superposition of nucleon--nucleon interactions. 

Thus, the background fluctuations contain contributions from independent particle
production and correlations corresponding to different physical origins. The background-subtracted fluctuations can be used in Eq.~(\ref{kTeqn}) to estimate \kT~with the knowledge of the temperature and volume from 
complementary analyses of hadron yields, calculated at the chemical freeze-out~\cite{Sharma:2018jqf,Adam:2015vda}. 

In addition to fluctuations  in the number of participant nucleons, several other processes
contribute to fluctuations of the charged particles multiplicity on an event-by-event basis~\cite{Mrowczynski:1997kz,Mrowczynski:2009wk}. These include long-range particle correlations, charge conservation, resonance production, radial flow, as well as  Bose-Einstein correlations. Since  these contributions can not be evaluated directly, the value of \kT~extracted and reported  in this work amounts to an upper limit.

The article is organized as follows. In section~II, the experimental setup and details of the data analysis method, including event selection, centrality selection, corrections for finite width of the centrality intervals, and particle losses are presented. In section~III, the measurements of the variances of multiplicity distributions are presented as a function of collision centrality. Additionally, the dependence of the fluctuations on the $\eta$ and $p_{\rm T}$ ranges of the measured charged hadrons are studied. The results are compared with calculations from selected event generators. In section~IV, methods used to estimate multiplicity
fluctuations resulting from the fluctuations of the number of participants are discussed.
An estimation of the isothermal compressibility for central collisions is made in section~V. 

\section{Experimental setup and analysis details}

The ALICE experiment~\cite{Aamodt:2008zz} is a multi-purpose detector designed to measure and identify particles produced in heavy-ion collisions at the LHC. The experiment consists of several central barrel detectors positioned inside a solenoidal magnet operated at 0.5~T~field parallel to the beam direction and a set of detectors placed at forward rapidities. The central barrel of the ALICE detector provides full azimuthal coverage for track reconstruction within a pseudorapidity ($\eta$) range of $|\eta|<0.8$.  
The Time Projection Chamber (TPC) is the main tracking detector of the central barrel, consisting of 159 pad rows grouped into 18 sectors that cover the full azimuth. The Inner Tracking System (ITS) consists of six layers of silicon detectors employing three different technologies. The two innermost layers are Silicon Pixel Detectors (SPD), followed by two layers of Silicon Drift Detectors (SDD), and finally, the two outermost layers are double-sided Silicon Strip Detectors (SSD). 
The V0 detector consists of two arrays of scintillators located on opposite sides of the interaction point (IP). It features full azimuthal coverage in the forward and backward rapidity ranges, $2.8 < \eta < 5.1$ (V0A) and $-3.7 < \eta <-1.7$ (V0C). The V0 detectors are used for event triggering purposes as well as to evaluate the collision centrality on an event-by-event basis~\cite{Abelev:2013qoq}.
The impact of
the detector response on the measurement of charged-particle multiplicity based on Monte
Carlo simulations is studied with the GEANT3 framework~\cite{Brun:1994aa}.

This analysis is based on Pb--Pb collision data recorded in 2010 at \snn~$=$~2.76~TeV with a minimum-bias trigger comprising of a combination of hits in the V0 detector and the two innermost (pixel) layers of the ITS. In total, 13.8 million minimum-bias events satisfy the event selection criteria.
The primary interaction vertex of a
collision is obtained by extending correlated hits in the two SPD layers to the beam axis.
The longitudinal position of the interaction vertex in the beam ($z$) direction ($V_{\rm z}$) is 
restricted to $|V_{\rm z}| < 10$~cm to ensure a uniform acceptance in the central $\eta$ region.  
The interaction vertex is also obtained from TPC tracks. 
The event selection includes an additional vertex selection criterion, where the difference between the vertex using TPC tracks and the vertex using the SPD is less than 5~mm in the $z$-direction. This selection criterion greatly suppresses the contamination of the primary tracks by secondary tracks resulting from weak decays and spurious interactions of particles within the apparatus.

Charged particles are reconstructed using the combined information of the TPC and ITS~\cite{Aamodt:2008zz}. 
In the TPC, tracks are reconstructed from a collection of space points (clusters). 
The selected tracks are required to have at least 80 reconstructed space points. 
Different combinations of tracks in the TPC and SPD hits are utilized to correct for detector acceptances 
and efficiency losses.  
To suppress contributions from secondary tracks (i.e., charged particles produced by weak decays and interactions of particles with materials of the detector), the analysis is restricted to 
charged-particle tracks featuring a distance of closest approach (DCA) to the interaction vertex,
DCA$_{\rm xy} < 2.4$~cm in the transverse plane and of DCA$_{\rm z} < 3.2$~cm along the beam direction. The tracks are additionally restricted to 
the kinematic range, $|\eta|<0.8$ and $0.2 < p_{\rm T} < 2.0$~GeV/$c$.

\subsection{Centrality selection and the effect of finite width of the centrality intervals}

The collision centrality is estimated based on the sum of the amplitudes of the V0A and V0C signals (known as the V0M collision centrality estimator)~\cite{Abelev:2013qoq}. Events are classified in percentiles of the hadronic cross section using this estimator. The 
average number of participants in a centrality class, denoted by $N_{\rm part}$,  is 
obtained by comparing the V0M multiplicity to a geometrical Glauber model~\cite{Miller:2007ri}. 
Thus, the centrality of the collision is
measured based on the V0M centrality estimator, whereas the measurement of multiplicity fluctuations is 
based on charged particles measured within the acceptance of the TPC.

A given centrality class is a collection of events of
measured multiplicity distributions within a range in V0M corresponding to a mean number
of participants, $\langle N_{\rm part} \rangle$.
This results in additional fluctuations
in the number of particles within each centrality class. To account for these
fluctuations, a centrality interval width correction is employed.
The procedure involves dividing a broad centrality class into several narrow intervals and 
correcting for the finite interval using weighted moments according to~\cite{Luo:2013bmi,Sahoo:2012wn},
\begin{equation}
X = \frac{\sum_{\rm i}  n_{\rm i}X_{\rm i}}{\sum_{\rm i}n_{\rm i}}.
\end{equation}
Here, the index $i$~runs over the narrow centrality intervals. $X_{\rm i}$ and $n_{\rm i}$ are the corresponding moments of the distribution and number of events in the $i^{\rm th}$ interval, respectively. With this, one obtains,   
$N = \sum_{\rm i}n_{\rm i}$ as the total number of events in the broad centrality interval.

The centrality resolution of the combined V0A and V0C signals ranges from 0.5\% in central to 2\% in the most peripheral collisions~\cite{Abelev:2013qoq}. A correction for the finite width of centrality intervals has been made with Eq.~4 
using 0.5\% centrality intervals
from central to $40\%$~cross section and $1\%$~intervals for the rest of the centrality classes.

\subsection{Efficiency correction}

The detector efficiency factors ($\varepsilon$) were evaluated in  bins of pseudorapidity $\eta$, azimuthal angle $\varphi$, and $p_{\rm T}$. 
By defining 
$N_{\rm ch}(x)$ as the number of produced particles in a phase-space bin at $x$, $n(x)$ as the number of observed particles at $x$, and $\varepsilon(x)$ as the detection efficiency, 
the first and second factorial moments of the multiplicity distributions can be corrected for particle losses according to the procedure outlined in Ref.~\cite{Bzdak:2013pha,Luo:2014rea}:
\begin{equation}
F_{\rm 1}= \langle N_{\rm ch} \rangle = \sum_{i=1}^m \langle N_{\rm ch}(x_{\rm i}) \rangle = \sum_{i=1}^m \frac{n(x_{\rm i})}{\varepsilon(x_{\rm i})},
\label{an-1}
\end{equation}
and
\begin{equation} 
F_{\rm 2} =  \sum_{i=1}^m \sum_{j=i}^m \frac{\langle n(x_{\rm i})(n(x_{\rm j})-\delta_{x_{\rm i}x_{\rm j}})\rangle}{\varepsilon(x_{\rm i}) \varepsilon(x_{\rm j})},
\label{an-2}
\end{equation} 
respectively. Here, $m$ denotes the index of the phase-space bins and $i,j$ are the bin indexes. 
$\delta_{x_{\rm i}x_{\rm j}}=1$ if $x_{\rm i}=x_{\rm j}$ and zero otherwise.  
The variance
of the charged-particle multiplicity is then calculated as:
\begin{equation}
\sigma_{\rm ch}^{\rm 2} = F_{\rm 2} + F_{\rm 1} - F_{\rm 1}^{\rm 2}.
\label{an-3}
\end{equation}

The correction procedure is validated by a Monte Carlo study employing two million Pb--Pb events at
$\sqrt{s_{\rm NN}}$~$=$~2.76~TeV generated using the HIJING event generator~\cite{Deng:2010xg}, and
passed through GEANT3 simulations of the experimental setup, taking care of the acceptances of the detectors. 
The efficiency dependencies on $\eta$, $\varphi$, and 
$p_{\rm T}$ are calculated from the ratio of the number of reconstructed charged particles by the number of produced particles. In order to account for the $p_{\rm T}$ dependence of efficiency,  the full $p_{\rm T}$ range ($0.2 < p_{\rm T} < 2.0$~GeV/$c$) was divided to nine bins
(0.2-0.3, 0.3-0.4, 0.4-0.5, 0.5-0.6, 0.6-0.8, 0.8-1.0, 1.0-1.2, 1.2-1.6, 1.6-2.0) with larger number of bins in low $p_{\rm T}$ ranges. In the Monte Carlo closure test, 
the values of $\langle N_{\rm ch} \rangle$, $\sigma_{\rm ch}$, and $\omega_{\rm ch}$ of
the efficiency corrected results from the simulated events are compared to those of HIJING at the generator level to obtain the corrections.
By construction, the efficiency corrected values for $\langle N_{\rm ch} \rangle$ match with those from the generator, whereas $\sigma_{\rm ch}$ and $\omega_{\rm ch}$ values differ by $\sim$0.7\%  and  $\sim$1.4\%, respectively. These differences are included in the systematic uncertainties.

\subsection{Statistical and systematic uncertainties}

The statistical uncertainties of the moments of multiplicity distributions are calculated based on the method of error propagation derived from the delta theorem~\cite{Luo:2011tp}. The systematic uncertainties have been evaluated by considering the effects of various criteria
in track selection, vertex determination, and efficiency corrections.

The systematic uncertainties related to the track selection criteria were obtained by varying the
track reconstruction method and track quality cuts.
The nominal analysis was carried out with charged particles reconstructed within the TPC and ITS.
For systematic checks, the full analysis is repeated for tracks reconstructed using only the TPC information. 
The differences in the values of $\langle N_{\rm ch} \rangle$, $\sigma_{\rm ch}$, and $\omega_{\rm ch}$ resulting from the track selections using the two methods are listed in Table~\ref{table:listing1} as a part of the systematic uncertainties. The $DCA_{\rm xy}$ and $DCA_{\rm z}$ of the tracks are varied by
 $\pm 25$\% to obtain 
the systematic uncertainties caused by variations in the track quality selections.
The effect of the selection of events based on the vertex position is studied 
by restricting the $z$-position of the vertex to $\pm 5$~cm from the nominal $\pm 10$~cm, and  
additionally by removing restrictions on $V_{\rm x}$ and $V_{\rm y}$. 
The efficiency correction introduces additional systematic uncertainty as discussed earlier.
The experimental data were recorded for two different magnetic field polarities. The two data sets are analyzed separately and the differences are taken as a source of systematic uncertainties. 

The individual sources of systematic uncertainties discussed above are considered uncorrelated and summed in quadrature to obtain the total systematic errors reported in this work. 
 Table~\ref{table:listing1} lists the systematic uncertainties associated with the values of $\langle N_{\rm ch} \rangle$, $\sigma_{\rm ch}$, and $\omega_{\rm ch}$.

\begin{table}[tbp]
\centering 
\caption{Systematic uncertainties on the mean, standard deviation, and scaled variance of charged-particle multiplicity distributions from different sources. The ranges of uncertainties quoted correspond to central to peripheral collisions.}
\begin{tabular}{|m{10.0em} || m{2.5cm} | m{2.5cm} | m{3.4cm} |}
\hline
Source    & $\langle N_{\rm ch} \rangle$ & $\sigma_{\rm ch}$ & $\omega_{\rm ch}$\\
\hline
Track selection & $3.5-4.8$\% & $3.8-6.0$\% & $4.0-7.5$\%\\
\hline
Variation of $DCA_{\rm xy}$  & $0.5-0.9\%$ & $0.8-1.2\%$ &
$1.3-1.6\%$\\
\hline
Variation of $DCA_{\rm z}$  & $0.4-0.9\%$ & $0.7-1.0\%$ &
$1.2-1.7\%$\\
\hline
Vertex ($V_{\rm z}$) selection & $0.1-0.5\%$ & 0.5\% & $0.1-0.8$\%\\
\hline
Removal of $V_{\rm x}, V_{\rm y}$ selections & 0.1\% &
0.2\% & 0.5\% \\
\hline
Efficiency correction & $<$0.1\% & 0.7\% & 1.4\% \\
\hline
Magnetic polarity & $0.2-1.0$\% & $0.5-1.5$\% &
$0.8-1.7$\%\\
\hline
\hline
Total & $3.5-5.1$\% & $4.1-6.4$\% & $4.8-8.3$\% \\
\hline
\end{tabular} 
\label{table:listing1}
\end{table}

\section{Results and discussions}

\begin{figure}[tbp]
\begin{center}
 \includegraphics[scale=0.5]{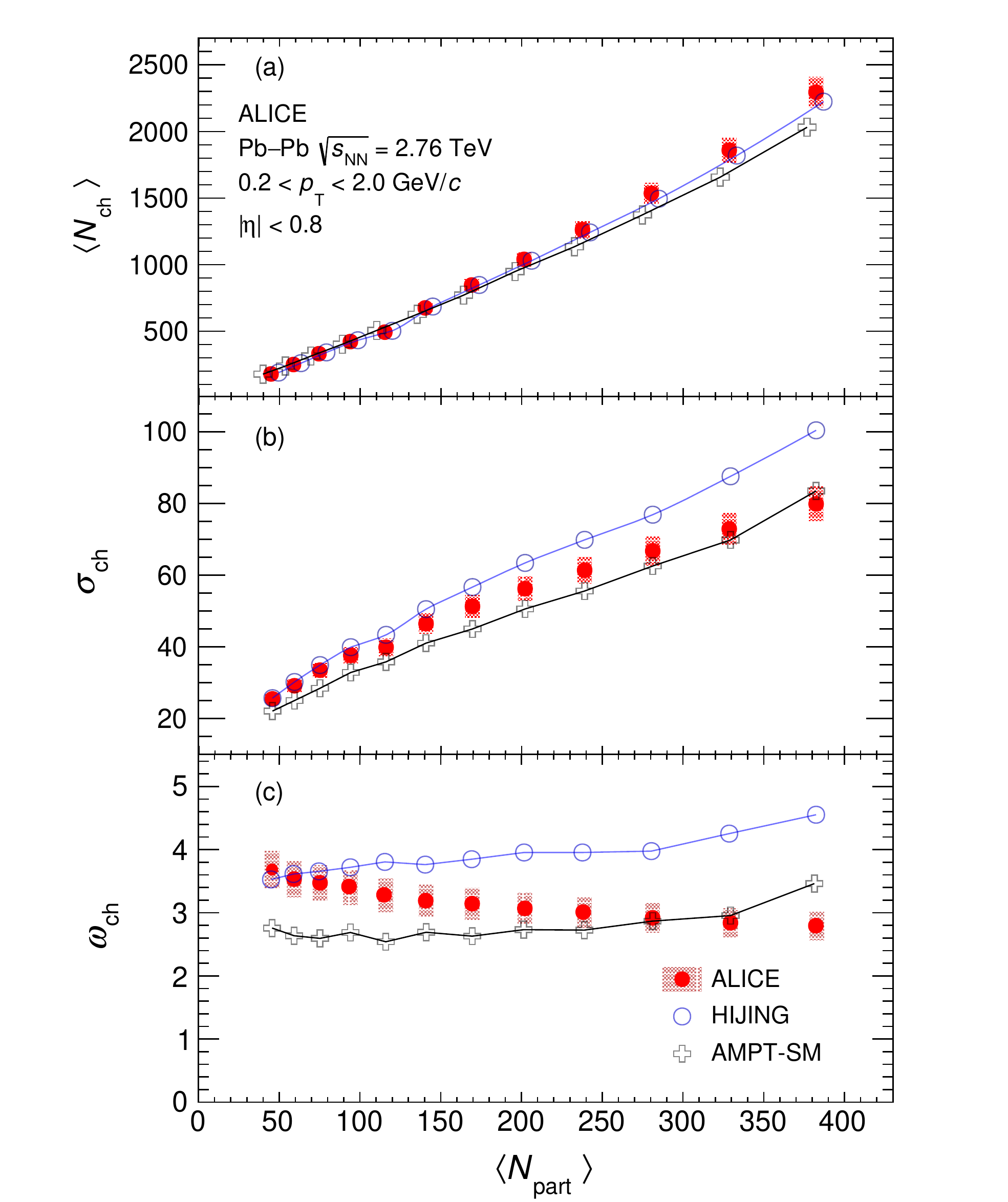}
\caption{Mean ($\langle N_{\rm ch} \rangle$), standard deviation ($\sigma_{\rm ch}$), and scaled variance ($\omega_{\rm ch}$) of charged-particle multiplicity distributions as a function of the number of participating nucleons for experimental data along with
HIJING and AMPT (string melting) models for Pb--Pb collisions at \snn~$=$~2.76~TeV, shown in panels (a), (b), and (c), respectively. For panel (a),  $\langle N_{\rm part} \rangle$ for the two models are shifted for better visibility.
The statistical uncertainties are smaller than the size of the markers. 
The systematic uncertainties are presented as filled boxes. 
}
\label{fig systematicall}
\end{center}
\end{figure}

Figure~\ref{fig systematicall} shows the corrected mean ($\langle N_{\rm ch} \rangle$), standard deviation ($\sigma_{\rm ch}$), and scaled variance ($\omega_{\rm ch}$) as a function of $\langle N_{\rm part} \rangle$ for the centrality range considered (0-60\%) corresponding to  $N_{\rm part}>45$. 
Uncertainties on  the estimated  number of participants, $\langle N_{\rm part} \rangle$, obtained from Ref. ~\cite{Aamodt:2008zz},  are  smaller than the width of the solid red circles used to present the  data in the centrality range considered in this measurement. 
It is observed that the values of $\langle N_{\rm ch} \rangle$ and 
$\sigma_{\rm ch}$ increase with increasing $\langle N_{\rm part} \rangle$.
The value of $\omega_{\rm ch}$ 
decreases monotonically by $\sim$29\% from peripheral to central collisions.

\begin{figure}[h]
\begin{center}
\includegraphics[scale=0.5]{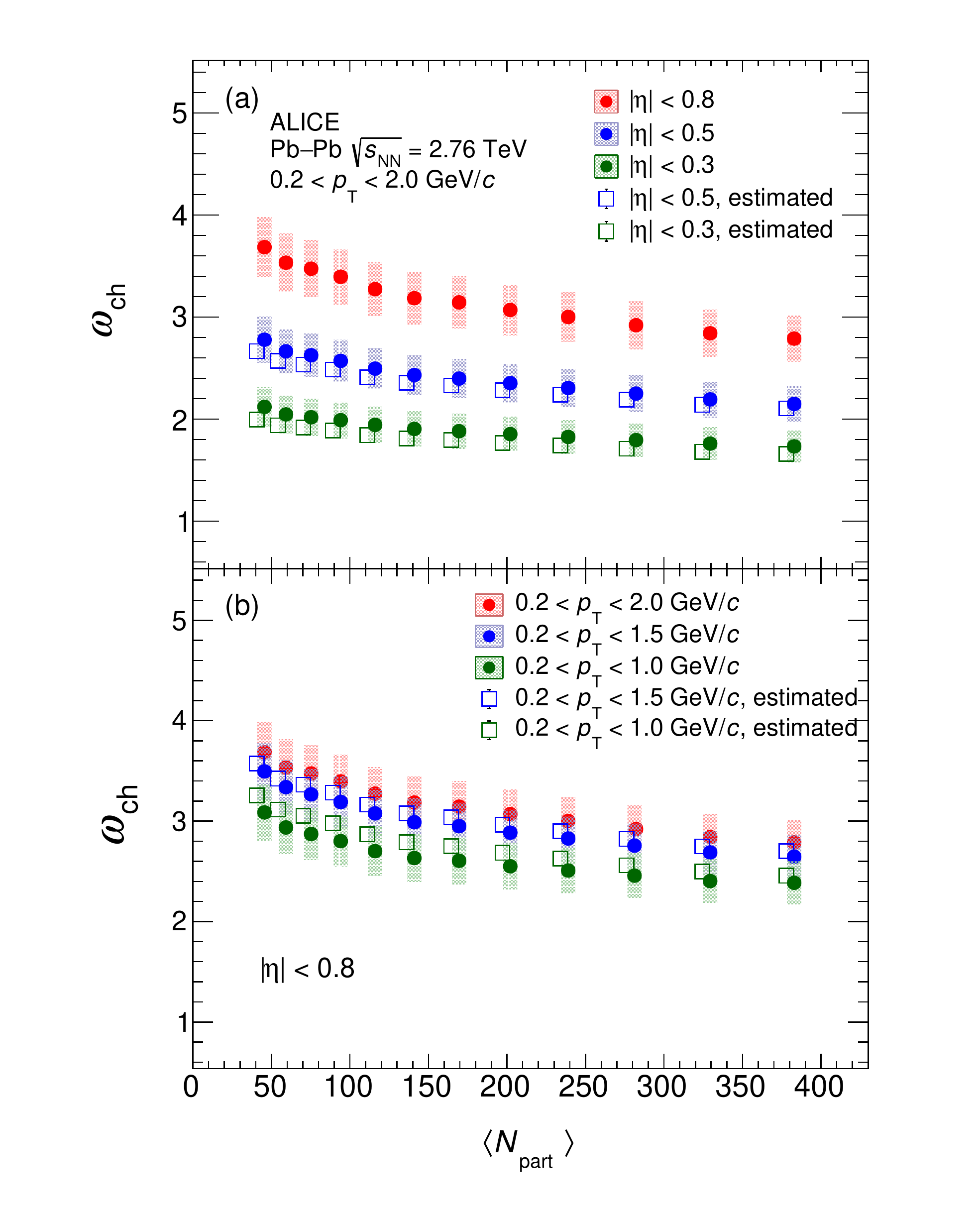}
\caption{
Scaled variances of charged-particle multiplicity distributions 
for different $\eta$ and 
 \pT~ranges as a function of number of participating nucleons 
measured in Pb--Pb collisions at \snn~$=$~2.76~TeV, shown in
panels (a), and (b), respectively.
The estimated  $\omega_{\rm ch}$ for $|\eta|<0.3$ and 
$|\eta|<0.5$ are obtained from the experimental data of $|\eta|<0.8$ by using Eq.~\ref {empirical}. The 
estimated $\omega_{\rm ch}$ for $0.2 < p_{\rm T} < 1.5$~GeV/$c$ and $0.2 < p_{\rm T} < 1.0$~GeV/$c$ 
are obtained from the experimental data of $0.2 < p_{\rm T} < 2.0$~GeV/$c$, also by using Eq.~\ref {empirical}. 
The statistical uncertainties are smaller than the size of the markers. 
The systematic uncertainties are presented as filled boxes. 
}
\label{acceffect}
\end{center}
\end{figure}

\subsection{Comparison with models}

The measured $\omega_{\rm ch}$ values are compared with the results of simulations with the HIJING and the string melting option of the AMPT models.
HIJING~\cite{Deng:2010xg} is a Monte Carlo event generator for parton and particle production in high-energy hadronic and nuclear collisions and is based on QCD-inspired models which incorporate mechanisms such as multiple minijet production, soft excitation, nuclear shadowing of parton distribution functions, and jet interactions in the dense hadronic matter.
The HIJING model treats a nucleus--nucleus collision as a superposition of many binary nucleon--nucleon collisions. 
In the AMPT model~\cite{Lin:2004en}, the initial parton momentum distribution is generated from the
HIJING model. In the default mode of AMPT, energetic partons recombine and hadrons are produced via string fragmentation. The string melting mode of the model includes a fully partonic phase that hadronises through quark coalescence.

In order to enable a proper comparison with data obtained in this work, Monte Carlo events produced with HIJING and AMPT  are grouped in collision centrality classes based on generator level charged-particle multiplicities computed in the ranges $2.8 < \eta < 5.1$ and $-3.7 < \eta <-1.7$,  
corresponding to the V0A and V0C pseudorapidity coverages.
The results of the scaled variances from the two event generators 
are presented in Fig.~\ref{fig systematicall} as a function of the estimated number of 
participants, $N_{\rm part}$.
As a function of increasing centrality, 
the $\omega_{\rm ch}$ values obtained from the event generators show upward trends,
which are opposite to those of the experimental data.
It is to be noted that the Monte Carlo event generators are successful in reproducing
the mean of multiplicity distributions. This follows from the fact that the particle multiplicities are
proportional to the cross sections. On the other hand, the widths of the distributions originate from 
fluctuations and correlations associated with effects of different origins, such as
long-range correlations, Bose--Einstein correlations, resonance decays, and charge conservation.
Because of this, the event generators fall short of reproducing the observed scaled variances.

\subsection{Scaled variance dependence on pseudorapidity acceptance and \pT~range}

Charged-particle multiplicity distributions depend on the acceptance of the detection region. Starting with the measured multiplicity fluctuations within  
$|\eta|<0.8$ and $0.2<p_{\rm T}<2.0$~GeV/$c$
with a mean $\langle N_{\rm ch} \rangle$ and scaled variance of $\omega_{\rm ch}$, 
the scaled variance ($\omega_{\rm ch}^{\rm acc}$) for a fractional acceptance in $\eta$ or for a limited  \pT~range with mean of $\langle N_{\rm ch}^{\rm acc} \rangle$ can be expressed as~\cite{Adare:2008ns},
\begin{equation}
\omega_{\rm ch}^{\rm acc} = 1 + f^{\rm acc} (\omega_{\rm ch}-1),
\label{empirical}
\end{equation}
\begin{equation}
{\rm where}~~ 
f^{\rm acc} = \frac{\langle N_{\rm ch}^{\rm acc} \rangle}{\langle N_{\rm ch} \rangle}.
\end{equation}
This empirical estimation for the acceptance dependence of the scaled variance is valid assuming
that there are no significant correlations present over the acceptance range being studied. 
The validity of this dependence has been checked by comparing the experimental data of scaled variances at 
reduced acceptances along with the results from the above calculations. 
This is shown in Fig.~\ref{acceffect} for different $\eta$ or \pT~ranges.  
In the top panel, the scaled variances are shown, as a function of $\langle N_{\rm part} \rangle$, for
three $\eta$ ranges.
The solid symbols show the results of measured scaled variances, whereas open symbols show the estimated values for the two reduced $\eta$ windows. The calculated values yield a good description of the measured data points. The choice of the \pT~range also affects the multiplicity of an event. 
In the bottom panel of Fig.~\ref{acceffect}, the scaled variances are shown, as a function of $\langle N_{\rm part} \rangle$,
for three \pT~ranges keeping $|\eta|<0.8$. A decrease in the value of $\omega_{\rm ch}$ is observed with the decrease of the $p_{\rm T}$ window. The results from the calculations of scaled variances are  compared to the measured data points. The calculated values are close to those of the measurement. 
This estimation of the scaled variances of multiplicity distributions 
is particularly useful in extrapolating fluctuations to different coverages.

\begin{figure}[tbp]
\begin{center}
\includegraphics[scale=0.5]{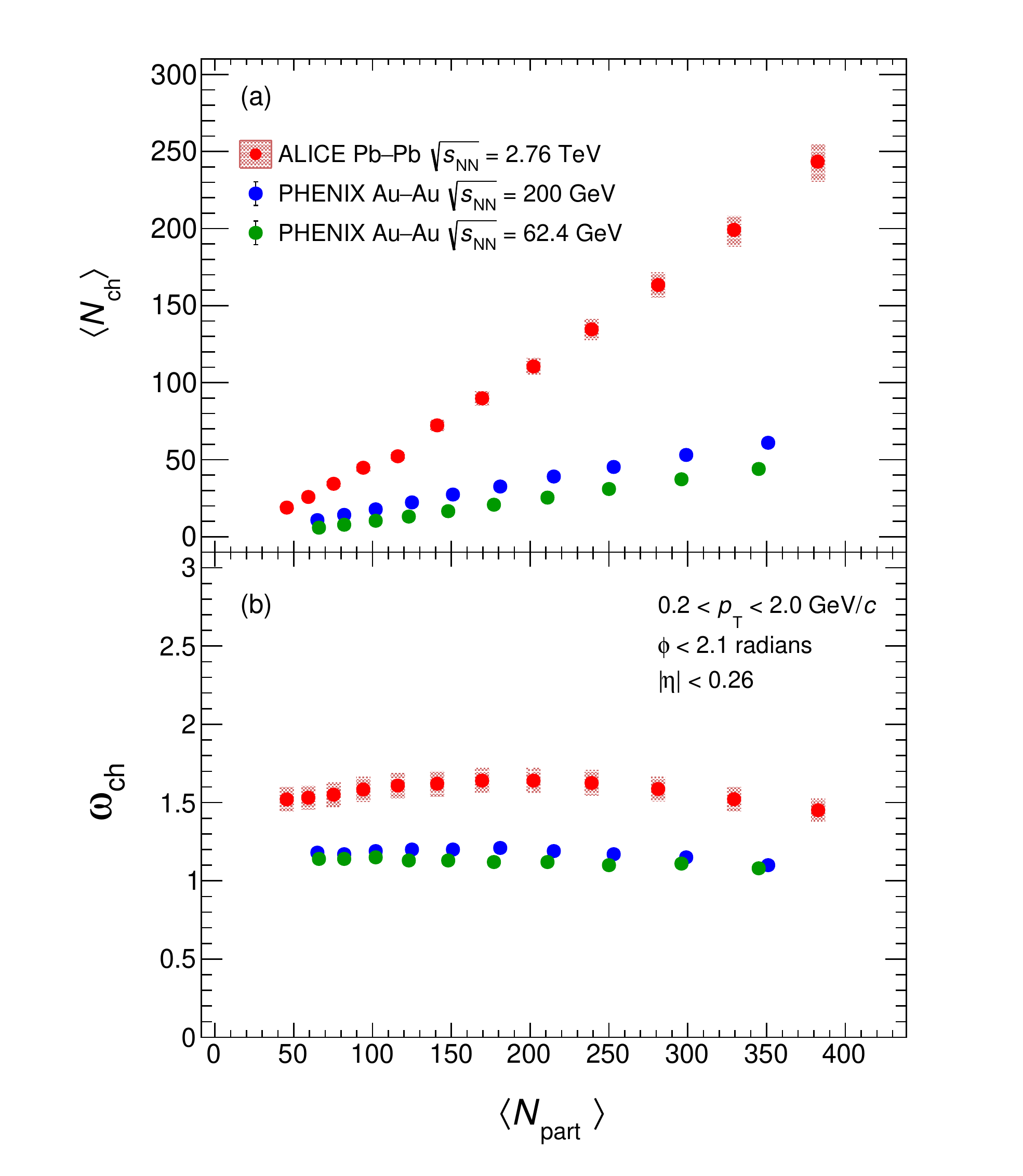}
\includegraphics[scale=0.5]{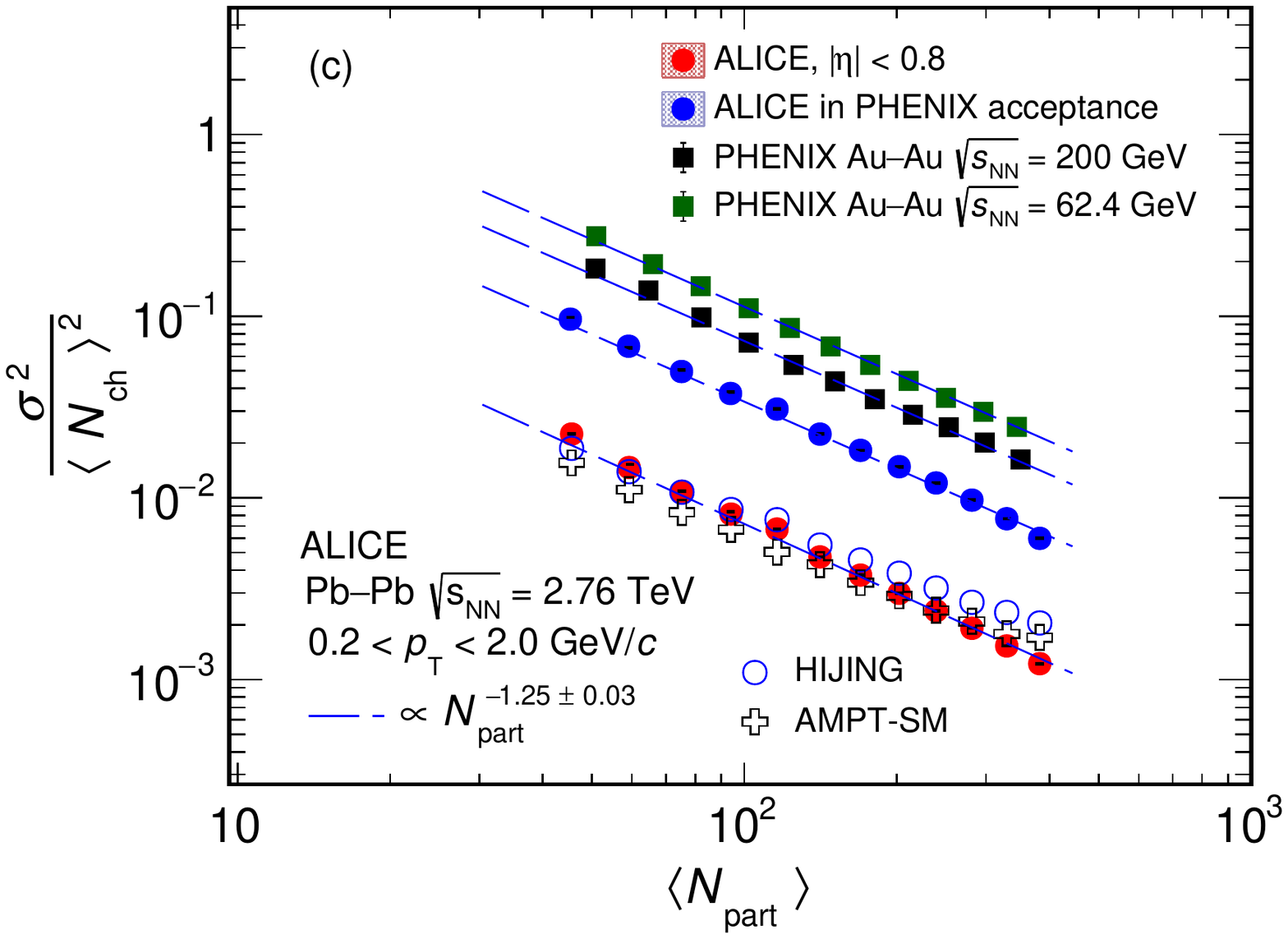}
\caption{Comparison of $\langle N_{\rm ch} \rangle$, $\omega_{\rm ch}$, and
$\sigma_{\rm ch}^2/\langle N_{\rm ch} \rangle^2$ measured in this work
based on the acceptance of the PHENIX experiment with results reported by 
PHENIX~\cite{Adare:2008ns} as a function of number of participating nucleons, shown in panels
(a), (b), and (c), respectively.
The statistical uncertainties are smaller than the size of the markers. 
The systematic uncertainties are presented as filled boxes.
}
\label{lowenergy}
\end{center}
\end{figure}

\subsection{Comparison to scaled variances at lower collision energies}

Scaled variances of charged-particle multiplicity distributions were 
earlier reported by the  PHENIX Collaboration at RHIC for Au--Au collisions at \snn~=~62.4~GeV and 200~GeV~\cite{Adare:2008ns}. 
The beam--beam counters (BBC) in PHENIX covering the 
full azimuthal angle in the pseudorapidity range $3.0 < |\eta| < 3.9$ 
provided the minimum-bias trigger and were used for centrality selection.
The pseudorapidity acceptance of the PHENIX experiment amounted to 
$|\eta|<0.26$ with an effective average azimuthal active area of 2.1 radian
and $0.2<p_{\rm T}<2.0$~GeV/$c$ for charged particle measurements.
The published results of mean and scaled variances of charged-particles 
were corrected for fluctuations of the collision geometry within a centrality bin. 
This was performed by comparing fluctuations from simulated HIJING events with a 
fixed impact parameter to events with a range of impact parameters covering the width of 
the centrality bin, as determined from Glauber model simulations. 
The corrected results are reproduced in Fig.~\ref{lowenergy} for
the two collision energies.
To enable an appropriate comparison with results reported by PHENIX, 
the ALICE data are reanalyzed by imposing the same kinematic ranges as in PHENIX, and 
the resulting mean and scaled variances are presented in Fig.~\ref{lowenergy}.
It is observed that for the same acceptance and kinematic cuts, 
the mean values and the scaled variances are larger at the LHC energy compared to those obtained at RHIC energies.

It is also of interest to study 
$\frac{\sigma_{\rm ch}^{\rm 2}}{{\langle N_{\rm ch} \rangle}^{\rm 2}}$, 
the ratio of the variance by the square of the average multiplicity
as a function of collision centrality. At lower beam energies, these distributions obey a power-law relative to the number of participants~\cite{Mitchell:2007wz}. In
the lower panel of Fig.~\ref{lowenergy}, 
the values of
 $\frac{\sigma_{\rm ch}^{\rm 2}}{{\langle N_{\rm ch} \rangle}^{\rm 2}}$ are
 presented 
as a function of $\langle N_{\rm part} \rangle$ for the ALICE data, for the
common coverage of ALICE and PHENIX data, as well as PHENIX data at two collision energies. 
The data points are fitted by a scaling curve, 
$\frac{\sigma_{\rm ch}^{\rm 2}}{{\langle N_{\rm ch} \rangle}^{\rm 2}} = A \cdot N_{\rm part}^{\alpha}$.
The exponent $\alpha = -\rm 1.25 \pm 0.03$ fits the four sets of experimental data well with 
$\chi^2/{\rm ndf}$ (where $\rm {ndf}$ is the number of degrees of freedom) as
0.88, 1.1, 0.95, 0.84
for
Pb--Pb collisions at $\sqrt{s_{\rm NN}}~=$~2.76~TeV with the ALICE acceptance, the PHENIX
detector acceptance and Au--Au collisions at 
$\sqrt{s_{\rm NN}}~=~$200~GeV and 62.4~GeV, respectively.
The scaling, first described by the PHENIX Collaboration~\cite{Mitchell:2007wz}, also holds for the ALICE data. The corresponding values of $\frac{\sigma_{\rm ch}^{\rm 2}}{{\langle N_{\rm ch} \rangle}^{\rm 2}}$ for HIJING and AMPT models for Pb–Pb collisions at $\sqrt{s_{\rm NN}}~=$~2.76~TeV are also displayed in Fig.~\ref{lowenergy}. 
The trends as a function of centrality are observed to be similar to those of the experimental data.
Fits with a similar scaling curve yield 
power-law exponents as $-1.1$ and $-1.05$ for  HIJING and AMPT models, respectively. These exponents for the models are lower compared those of the experimental data.

\section{Background to the measured multiplicity fluctuations}

The background to the measured multiplicity fluctuations contains contributions from
several sources. In this section, the background fluctuations are presented 
first from a participant model calculation and then the expectations from a Poisson distribution of particle multiplicity
are discussed.

In the wounded nucleon model, 
nucleus--nucleus (such as Pb--Pb) collisions are considered to be a superposition of individual nucleon--nucleon 
interactions. In this context, the 
fluctuations in multiplicity within a given centrality window arise in part from 
fluctuations in $N_{\rm part}$
and from fluctuations in the number of particles ($n$) produced by each
nucleon--nucleon interaction~\cite{Heiselberg:2000fk,Aggarwal:2001aa,Adare:2008ns}.  
The values of $n$ and
their fluctuations are also strongly dependent on the acceptance of the detector. 
Within the context of this  framework, the scaled variance of the background, 
$\omega_{\rm ch}^{\rm back}$, amounts to
\begin{equation}
    \omega_{\rm ch}^{\rm back} = \omega_{\rm n} + \langle n \rangle \omega_{N_{\rm part}},
\label{kTeqn2}
\end{equation}
where 
$\langle n \rangle$ is the average number of particles produced by each nucleon--source within the detector acceptance, $\omega_{\rm n}$ is the scaled variance of the fluctuations in $n$, and 
$\omega_{N_{\rm part}}$ denotes the fluctuations in $N_{\rm part}$. The variance,
$\omega_{N_{\rm part}}$ is calculated using event-by-event $N_{\rm part}$ from
the HIJING model. The distribution of \Npart~corresponds to the centrality obtained within the
V0 detector coverage ($2.8 < \eta < 5.1$ and $-3.7 < \eta <-1.7$).
The extracted values of $\omega_{N_{\rm part}}$
are corrected for the effects of the finite width of the centrality intervals.

\begin{figure}[tbp]
\begin{center}
 \includegraphics[scale=0.5]{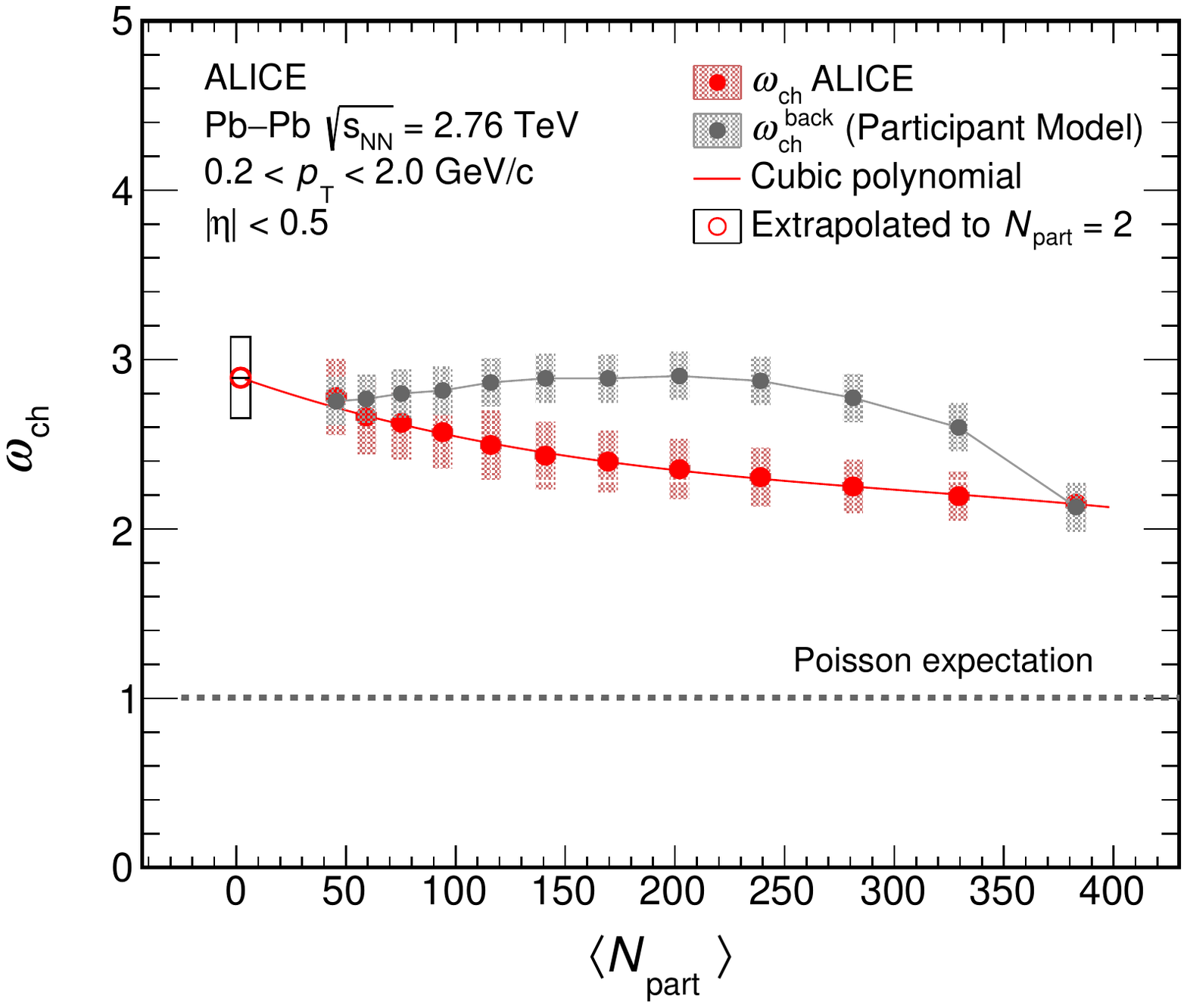}
\caption{
Scaled variance as a function of \aveNpart~for charged-particle multiplicity distributions and 
background fluctuations ($\omega_{\rm ch}^{\rm back}$) based on a
participant model calculation for  $|\eta| < 0.5$.
The expectation from Poisson-like particle production is indicated by the dotted line. 
The statistical uncertainties are smaller than the size of the markers. 
The systematic uncertainties are presented as filled boxes. 
}
\label{omega_background}
\end{center}
\end{figure}

For the central rapidity range ($|\eta| < 0.5$), 
the measured number of charged particles produced in
pp collisions within $0.2 < p_{\rm T} < 2.0$~GeV/$c$ at
$\sqrt{s}~$=$~2.76$~TeV~\cite{Acharya:2018qsh} yields
$\langle n \rangle=1.45 \pm 0.07$,
which is half of the measured value.
In order to calculate $\omega_{\rm n}$, 
an extrapolation of the measured $\omega_{\rm ch}$ is made to $N_{\rm part}=2$ using a polynomial fit function 
of the form $a+b x + c x^2 + d x^3$, which is shown in Fig.~\ref{omega_background}. In order to calculate $\omega_{\rm n}$, an extrapolation of the measured $\omega_{\rm ch}$ is made to $N_{\rm part}=2$ using a polynomial fit function of the form $a+b x + c x^2 + d x^3$, which is shown in Fig.~\ref{omega_background}. Since both the nucleon sources contributing to $N_{\rm part}=2$ are correlated, $\omega_{\rm n}$ becomes half of the extrapolated value, yielding
$\omega_{\rm n}=1.445\pm0.12$.
This result is also consistent with the value of 
$\omega_{\rm n} = 1.51\pm0.16$ obtained from the parameterization given by the PHENIX Collaboration~\cite{Adare:2008ns}.

Using the above numbers, $\omega_{\rm ch}^{\rm back}$ are calculated and 
plotted as a function of \aveNpart~as in Fig.~\ref{omega_background}.
The obtained trend in $\omega_{\rm ch}^{\rm back}$
mainly arises from the
centrality dependence of $\omega_{N_{\rm part}}$.
For most central collisions, the difference between
the measured and background $\omega_{\rm ch}$ is ~$0.02 \pm 0.18$,~which
is consistent with  zero within the uncertainties. 
Except for most central collisions, 
$\omega_{\rm ch}^{\rm back}$ is observed to be larger than $\omega_{\rm ch}$. 
Thus, it seems likely that the background estimated in this way from the participant model is overestimated.

For an ideal gas, the number fluctuations are described by the Poisson distribution. So, 
if the emitted particles are uncorrelated, then the 
multiplicity distributions become Poissonian, the magnitude of $\omega_{\rm ch}$ reduces
to unity,
which is independent of the multiplicity and thus independent of the centrality of the collision. As seen from
Fig.~\ref{omega_background}, the observed multiplicity fluctuations are significantly above the Poisson expectation for all centralities.

\section{Estimation of isothermal compressibility}
\label{section5}

Equation~(\ref{kTeqn})  relates the magnitude of the charged-particle multiplicity fluctuations 
to the isothermal compressibility. 
The calculation of \kT~requires knowledge of the  temperature and volume of the system. 
After the collision, as the system cools down, the hadronic yields are fixed
when the rate of inelastic collisions becomes
negligible (chemical freeze-out), but the transverse-momentum
distributions continue to change until elastic interactions also cease (kinetic freeze-out). 
The number of charged particles gets fixed at the time of chemical freeze-out
(except for long-lived resonances).  
As the calculation of \kT~depends on the fluctuations in the number of particles, 
the chemical freeze-out conditions are considered as input. 
The ALICE Collaboration has published the identified particle yields of  pions,
kaons, protons, light nuclei, and resonances~\cite{Abelev:2013vea,Adam:2015vda,Acharya:2017bso}.
The statistical hadronization models have been successful
in describing these yields and their ratios~\cite{Andronic:2017pug,Sharma:2018jqf,Petran:2013lja},
using temperature and volume as parameters at the chemical freeze-out. 
For most central Pb--Pb collisions at \snn~$=$~2.76~TeV, the ALICE data 
on yields of particles
in one unit of rapidity at midrapidity are in good agreement with 
$0.156\pm 0.002$~GeV and $5330\pm 505$~~fm$^3$, for temperature and volume, respectively~\cite{Acharya:2017bso}.
In addition, the charged-particle multiplicity within 
$|\eta| < 0.5$ in this centrality range is 
$\langle N_{\rm ch} \rangle = 1410 \pm 47~(syst)$.

Here, an attempt is made to estimate \kT~for Pb--Pb collisions 
using the charged-particle multiplicity fluctuations along with the
temperature, volume, and mean number of charged particles from above.
The measured multiplicity fluctuation for central collisions is $\omega_{\rm ch} = 2.15 \pm 0.1$.
In the absence of any background where the full fluctuation is attributed to have a
thermal origin, one would obtain $k_{T} =  52.1\pm 5.81$ fm$^3$/GeV. 
As the measured $\omega_{\rm ch}$ contains background fluctuations from
different sources, this value of \kT can be only be considered as an absolute upper limit.

In the previous section, the 
background fluctuations have been estimated from the 
participant model calculation as shown in Fig.~\ref{omega_background}.
For central collisions, the value of the 
measured fluctuation above that of the participant model fluctuation is
$\omega_{\rm ch} = 0.02 \pm 0.18$. This leads to $k_{T} = 0.48  \pm 4.32 $ fm$^3$/GeV.
On the other hand, the background fluctuations from the participant model for other 
centralities are larger
compared to the measured ones making the background-subtracted fluctuations negative.
So it is not possible to obtain 
estimates of \kT~for these centrality ranges based on the present model of participant fluctuations.

The measured multiplicity fluctuations can be viewed as combinations of
correlated and uncorrelated fluctuations. If the particle production is completely
uncorrelated, the system effectively behaves as an ideal gas, and
the multiplicity distribution is expected to follow a Poisson
distribution ($\omega_{\rm ch}= 1$). 
For central collisions,
fluctuations above the Poisson estimation gives,
$\omega_{\rm ch}=1.15\pm 0.06$, which in turn implies a value of
$k_{T} = 27.9 \pm 3.18 $~fm$^3$/GeV.

It may be noted that other sources 
likely also contribute to the background of the measured multiplicity fluctuations. A
quantitative determination of these effects requires further studies and theoretical
modeling, which is beyond the scope of this work. In view of this, the
estimation of \kT~from the 
background-subtracted
event-by-event multiplicity fluctuation provides an upper limit of its value.

It is imperative to put the extracted values of \kT~in perspective with respect to that of normal
nuclear matter. The incompressibility constant of normal nuclear matter at pressure $P$,
expressed as $K_{0} = 9 (\partial P/\partial \rho)$ at zero temperature and normal
nuclear density, $\rho=\rho_0$, has been determined to be
$K_{0} = 240\pm 20$~MeV~\cite{Blaizot:1980tw,Danielewicz:2002pu,Stone:2014wza}. 
Using the relation, $k_{T} = (9/\rho K_{0}) $, one obtains
the isothermal compressibility of nuclear matter 
to be $k_{T} \simeq 234 \pm 20$~fm$^3$/GeV. 
This is consistent with the expectation that normal nuclear matter at low temperature is more compressible than the high temperature matter produced at LHC energies (as of Eq.~\ref{kTeqn}).
From the above estimation, the value of $k_{T} = 27.9 \pm 3.18 $~fm$^3$/GeV, which corresponds to multiplicity fluctuations 
above the Poisson expectation, serves as a conservative upper limit, and is even significantly below the normal nuclear matter at low temperature.

\section{Summary}

Measurements of event-by-event fluctuations of charged-particle 
multiplicities are reported as a function of centrality in Pb--Pb collisions at \snn~$=$~2.76~TeV. 
The mean, standard deviation, and scaled variances of charged-particle multiplicities are 
presented for $|\eta|<0.8$ and $0.2<p_{\rm T}<2.0$~GeV/$c$ as a function of centrality.
A monotonically decreasing trend for the scaled variance is
observed from peripheral to central collisions. Corresponding results
from  HIJING
and AMPT event generators show a mismatch with the experimental results. 
The scaled variance of the multiplicity decreases with the reduction of the $\eta$ 
acceptance of the detector as well as with the decrease of the $p_{\rm T}$ range.
The multiplicity fluctuations are compared to the results from lower beam energies as reported by the PHENIX experiment. For the same acceptance, the observed scaled variances 
at RHIC energies are smaller compared to those observed at the LHC.

As multiplicity fluctuations are related to the isothermal compressibility of the system, the measured fluctuations are used to estimate \kT~in central Pb--Pb collisions  at \snn~$=$~2.76~TeV. 
The multiplicity fluctuations above the Poisson expectation 
case yields 
$k_{T} = 27.9 \pm 3.18 $~fm$^3$/GeV, which may still contain contributions
from additional uncorrelated particle production as well as 
from several non-thermal sources as discussed in section~\ref{section5}. 
Proper modeling of background subtraction needs to be developed by
accounting for all possible contributions from different physics origins, which 
is beyond the scope of the present work.
This result serves as a conservative upper limit of \kT~until
various contributions to the background are properly understood and evaluated.
The estimation of \kT~at lower collision energies and 
for different system-sizes is an interesting way to explore the QCD phase diagram from thermodynamics point of view.


\newenvironment{acknowledgement}{\relax}{\relax}
\begin{acknowledgement}
\section*{Acknowledgements}

The ALICE Collaboration would like to thank all its engineers and technicians for their invaluable contributions to the construction of the experiment and the CERN accelerator teams for the outstanding performance of the LHC complex.
The ALICE Collaboration gratefully acknowledges the resources and support provided by all Grid centres and the Worldwide LHC Computing Grid (WLCG) collaboration.
The ALICE Collaboration acknowledges the following funding agencies for their support in building and running the ALICE detector:
A. I. Alikhanyan National Science Laboratory (Yerevan Physics Institute) Foundation (ANSL), State Committee of Science and World Federation of Scientists (WFS), Armenia;
Austrian Academy of Sciences, Austrian Science Fund (FWF): [M 2467-N36] and Nationalstiftung f\"{u}r Forschung, Technologie und Entwicklung, Austria;
Ministry of Communications and High Technologies, National Nuclear Research Center, Azerbaijan;
Conselho Nacional de Desenvolvimento Cient\'{\i}fico e Tecnol\'{o}gico (CNPq), Financiadora de Estudos e Projetos (Finep), Funda\c{c}\~{a}o de Amparo \`{a} Pesquisa do Estado de S\~{a}o Paulo (FAPESP) and Universidade Federal do Rio Grande do Sul (UFRGS), Brazil;
Ministry of Education of China (MOEC) , Ministry of Science \& Technology of China (MSTC) and National Natural Science Foundation of China (NSFC), China;
Ministry of Science and Education and Croatian Science Foundation, Croatia;
Centro de Aplicaciones Tecnol\'{o}gicas y Desarrollo Nuclear (CEADEN), Cubaenerg\'{\i}a, Cuba;
Ministry of Education, Youth and Sports of the Czech Republic, Czech Republic;
The Danish Council for Independent Research | Natural Sciences, the VILLUM FONDEN and Danish National Research Foundation (DNRF), Denmark;
Helsinki Institute of Physics (HIP), Finland;
Commissariat \`{a} l'Energie Atomique (CEA) and Institut National de Physique Nucl\'{e}aire et de Physique des Particules (IN2P3) and Centre National de la Recherche Scientifique (CNRS), France;
Bundesministerium f\"{u}r Bildung und Forschung (BMBF) and GSI Helmholtzzentrum f\"{u}r Schwerionenforschung GmbH, Germany;
General Secretariat for Research and Technology, Ministry of Education, Research and Religions, Greece;
National Research, Development and Innovation Office, Hungary;
Department of Atomic Energy Government of India (DAE), Department of Science and Technology, Government of India (DST), University Grants Commission, Government of India (UGC) and Council of Scientific and Industrial Research (CSIR), India;
Indonesian Institute of Science, Indonesia;
Istituto Nazionale di Fisica Nucleare (INFN), Italy;
Institute for Innovative Science and Technology , Nagasaki Institute of Applied Science (IIST), Japanese Ministry of Education, Culture, Sports, Science and Technology (MEXT) and Japan Society for the Promotion of Science (JSPS) KAKENHI, Japan;
Consejo Nacional de Ciencia (CONACYT) y Tecnolog\'{i}a, through Fondo de Cooperaci\'{o}n Internacional en Ciencia y Tecnolog\'{i}a (FONCICYT) and Direcci\'{o}n General de Asuntos del Personal Academico (DGAPA), Mexico;
Nederlandse Organisatie voor Wetenschappelijk Onderzoek (NWO), Netherlands;
The Research Council of Norway, Norway;
Commission on Science and Technology for Sustainable Development in the South (COMSATS), Pakistan;
Pontificia Universidad Cat\'{o}lica del Per\'{u}, Peru;
Ministry of Education and Science, National Science Centre and WUT ID-UB, Poland;
Korea Institute of Science and Technology Information and National Research Foundation of Korea (NRF), Republic of Korea;
Ministry of Education and Scientific Research, Institute of Atomic Physics and Ministry of Research and Innovation and Institute of Atomic Physics, Romania;
Joint Institute for Nuclear Research (JINR), Ministry of Education and Science of the Russian Federation, National Research Centre Kurchatov Institute, Russian Science Foundation and Russian Foundation for Basic Research, Russia;
Ministry of Education, Science, Research and Sport of the Slovak Republic, Slovakia;
National Research Foundation of South Africa, South Africa;
Swedish Research Council (VR) and Knut \& Alice Wallenberg Foundation (KAW), Sweden;
European Organization for Nuclear Research, Switzerland;
Suranaree University of Technology (SUT), National Science and Technology Development Agency (NSDTA) and Office of the Higher Education Commission under NRU project of Thailand, Thailand;
Turkish Energy, Nuclear and Mineral Research Agency (TENMAK), Turkey;
National Academy of  Sciences of Ukraine, Ukraine;
Science and Technology Facilities Council (STFC), United Kingdom;
National Science Foundation of the United States of America (NSF) and United States Department of Energy, Office of Nuclear Physics (DOE NP), United States of America.
\end{acknowledgement}

\bibliographystyle{utphys}
\bibliography{ALICEMultFluc_Sumit} 

\newpage
\appendix

%
%

\section{The ALICE Collaboration}
\label{app:collab}

\small
\begin{flushleft} 

S.~Acharya$^{\rm 143}$, 
D.~Adamov\'{a}$^{\rm 98}$, 
A.~Adler$^{\rm 76}$, 
J.~Adolfsson$^{\rm 83}$, 
G.~Aglieri Rinella$^{\rm 35}$, 
M.~Agnello$^{\rm 31}$, 
N.~Agrawal$^{\rm 55}$, 
Z.~Ahammed$^{\rm 143}$, 
S.~Ahmad$^{\rm 16}$, 
S.U.~Ahn$^{\rm 78}$
% ...

\bigskip

\bigskip 

\textbf{\Large Affiliation Notes}

\bigskip 

% ...
$^{\rm V}$ Also at: Institute of Theoretical Physics, University of Wroclaw, Poland\\

\bigskip

\bigskip 

\textbf{\Large Collaboration Institutes}

\bigskip 

% ...
$^{16}$ Department of Physics, Aligarh Muslim University, Aligarh, India\\
% ...
$^{31}$ Dipartimento DISAT del Politecnico and Sezione INFN, Turin, Italy\\
% ...
$^{35}$ European Organization for Nuclear Research (CERN), Geneva, Switzerland\\
% ...
$^{55}$ INFN, Sezione di Bologna, Bologna, Italy\\
% ...
$^{76}$ Johann-Wolfgang-Goethe Universit\"{a}t Frankfurt Institut f\"{u}r Informatik, Fachbereich Informatik und Mathematik, Frankfurt, Germany\\
% ...
$^{78}$ Korea Institute of Science and Technology Information, Daejeon, Republic of Korea\\
% ...
$^{83}$ Lund University Department of Physics, Division of Particle Physics, Lund, Sweden\\
% ...
$^{98}$ Nuclear Physics Institute of the Czech Academy of Sciences, \v{R}e\v{z} u Prahy, Czech Republic\\
% ...
$^{143}$ Variable Energy Cyclotron Centre, Homi Bhabha National Institute, Kolkata, India\\
% ...

\end{flushleft} 
  
\end{document}